\PassOptionsToPackage{pdfpagelabels=false}{hyperref} 
\documentclass[usenatbib,useAMS]{mnras}

\newcommand\lsim{\mathrel{\rlap{\lower4pt\hbox{\hskip1pt$\sim$}}
        \raise1pt\hbox{$<$}}}
\newcommand\gsim{\mathrel{\rlap{\lower4pt\hbox{\hskip1pt$\sim$}}
        \raise1pt\hbox{$>$}}}

\usepackage{times}
\usepackage{graphicx}
\usepackage{verbatim}
\usepackage{tikz}
\usepackage{color}
\usepackage{amsbsy}
\usepackage{gensymb}
\usepackage{amsmath}
\usepackage{amssymb}

\usepackage[T1]{fontenc}
\usepackage{aecompl}
\usepackage{calligra}
\DeclareMathAlphabet{\mathcalligra}{T1}{calligra}{m}{n}
\DeclareFontShape{T1}{calligra}{m}{n}{<->s*[2.2]callig15}{}

\usepackage{hyperref}
\hypersetup{
    pdfnewwindow=true,      % links in new window
    colorlinks=true,       % false: boxed links; true: colored links
    linkcolor=blue,          % color of internal links
    citecolor=cyan,        % color of links to bibliography
    filecolor=blue,      % color of file links
    urlcolor=blue           % color of external links
}

\def\apj{ApJ}                 % Astrophysical Journal
\def\apjl{ApJL}               % Astrophysical Journal, Letters
\def\apjs{ApJS}               % Astrophysical Journal, Supplement
\def\mnras{MNRAS}             % Monthly Notices of the RAS
\def\aap{A\&A}                % Astronomy and Astrophysics
\def\nat{Nature}              % Nature
\def\araa{ARA\&A}             % Annual Review of Astronomy and Astrophysics
\def\Msun{{M_{\odot}}}

\def\bbeta{\boldsymbol{\beta}}
\def\btheta{\boldsymbol{\theta}}
\def\balpha{\boldsymbol{\alpha}}

\begin{document}

\title[Lensing from Eccentric SMBHBs]{Spikey: Self-Lensing Flares from Eccentric SMBH Binaries}
\author[B. X. Hu, D. J. D'Orazio, Z. Haiman, K. L. Smith, B. Snios, M. Charisi, R. Di Stefano]{Betty X. Hu$^{1,2}$\thanks{bhu@g.harvard.edu; daniel.dorazio@cfa.harvard.edu, zoltan@astro.columbia.edu},
  Daniel J. D'Orazio$^3$, Zolt\'an Haiman$^4$, Krista Lynne Smith$^5$,  
  \newauthor
  Bradford Snios$^6$, Maria Charisi$^7$, Rosanne Di Stefano$^6$\\
     $^1$Department of Physics, Harvard University, 17 Oxford Street, Cambridge, MA 02138, USA\\
     $^2$Department of Applied Physics and Applied Math, Columbia University, 500 West 120th Street, New York, NY 10027, USA\\
     $^3$Institute for Theory and Computation, Harvard University, 60 Garden Street, Cambridge, MA 02138, USA\\
     $^4$Department of Astronomy, Columbia University, 550 West 120th Street, New York, NY 10027, USA\\
     $^5$Stanford University KIPAC, SLAC, Menlo Park, CA 94025, USA\\
     $^6$Harvard-Smithsonian Center for Astrophysics, 60 Garden Street, Cambridge, MA 02138, USA\\
     $^7$TAPIR, California Institute of Technology, 1200 East California Blvd, Pasadena, CA 91125, USA}

\maketitle
\begin{abstract}
We examine the light curves of two quasars, motivated by recent suggestions that a supermassive black hole binary (SMBHB) can exhibit sharp lensing spikes. We model the variability of each light curve as due to a combination of two relativistic effects: the orbital relativistic Doppler boost and gravitational binary self-lensing. In order to model each system we extend previous Doppler plus self-lensing models to include eccentricity. The first quasar is identified in optical data as a binary candidate with a 20-yr period (Ark 120), and shows a prominent spike.  For this source, we rule out the lensing hypothesis and disfavor the Doppler-boost hypothesis due to discrepancies in the measured vs. recovered values of the binary mass and optical spectral slope.
The second source, which we nickname Spikey, is the rare case of an active galactic nucleus (AGN) identified in Kepler's high-quality, high-cadence photometric data.  For this source, we find a model, consisting of a combination of a Doppler modulation and a narrow symmetric lensing spike, that is consistent with an eccentric SMBHB with mass $M_{\text{tot}} = 3\times10^{7} \Msun$, rest-frame orbital period $T=418$ days, eccentricity $e=0.5$, and seen at an inclination of $8^{\circ}$ from edge-on. This interpretation can be tested by monitoring Spikey for periodic behavior and recurring flares in the next few years. In preparation for such monitoring we present the first X-ray observations of this object taken by the Neil Gehrels Swift observatory.
\end{abstract}


\maketitle

\section{\label{sec:intro}Introduction}

Motivated by the knowledge that supermassive black holes (SMBHs) are found in
the nuclei of most massive galaxies in the universe and that galaxies often
merge \citep{KormendyHo2013, richstone1998}, one expects supermassive black
hole binaries (SMBHBs) to be commonly found at the center of galaxies
\citep{begelman1980}. Despite this, very few SMBHBs have been detected,
typically at separations of several kpc
\citep[\textit{e.g.},][]{Dotti:2012:rev, Comerford:2013}. Recently, quasars
with periodically varying optical emission have been examined as candidates
for SMBHBs \citep{graham2015, charisi2016, LiuGezari+2019}. Such periodic
variability may be imprinted by the orbital motion of a sub-pc separation
SMBHB via time variable accretion
\citep[\textit{e.g.},][]{AL94,Hayasaki:2007,MM08, Cuadra:2009,
ShiKrolik:2012:ApJ, DHM:2013:MNRAS, Farris:2014, PG1302MNRAS:2015a,
ShiKrolik:2015, MunozLai:2016, Bowen+2019}, or by the relativistic Doppler
boost \citep{dorazio2015}. It has also been suggested that binary 
self-lensing, which would occur if the accretion flow from one black hole was
gravitationally lensed by its partner, could serve as a signature of SMBHBs
\citep[][hereafter D18]{dorazio2017}. Periodic self-lensing is expected in a
non-negligible number of SMBHBs if at least one black hole is accreting, and
is most dramatic in systems with small orbital inclinations relative to the
line of sight. The resulting lensing flare is symmetric and encodes the binary
orbital parameters; periodic repetition of such a flare would be a unique
signature of a sub-pc separation SMBHB.

In this paper, we extend the models of D18 to include orbital eccentricity. We
then use these models to analyze two SMBHB candidates by considering
periodically varying continuum emission caused by the relativistic Doppler
boost in addition to flares from periodic self-lensing. We choose to consider
continuum variability due to the Doppler boost instead of (or in addition to)
variable accretion because the former, without introducing extra model
parameters, provides a unique signature when combined with the lensing flares
(see D18 and \citealt{DoDi:2019}). Additionally, in the case of favorably
aligned systems, both effects are required to occur via relativity alone.

We save a rigorous search for self-lensing flares in existing time-domain
data for a future endeavor, and here consider two individually identified
Doppler + lensing candidates. One is the result of long term optical
monitoring, and the other a by-product of the Kepler mission. We find that our
Doppler + self-lensing model provides an excellent fit for the light curve of
an active galactic nucleus (AGN) identified in Kepler's data. We use our model
to predict when the putative next lensing flares will occur, providing us with
a clear test of the binary hypothesis for the Kepler AGN.

\section{\label{sec:cb}Candidate Binaries}

We first consider Arakelian 120 (hereafter Ark 120), a nearby radio-quiet type 1 AGN at a distance of 143 Mpc (z = $0.03271$). Spectroscopic and photometric data of this source exists from 1974 up to 2017. As reported in \cite{li2017}, long-term variations in the light curves of $V$-band flux densities, $5100\,\mathring{A}$ flux densities, and $H\beta$ integrated fluxes exhibit a sinusoidal pattern with a $\sim 20$ year period. In addition, the binned, merged light curve of $V\text{-band}$ and $5100\,\mathring{A}$ flux densities, shown in Figure \ref{fig:Ark120lc}, show two significant peaks around 1982 and 1997 (MJD $\sim$ 45,000 and 51,000, respectively), which we considered to be suggestive of binary self-lensing. 
We note that while this system has been put forward as a SMBHB candidate based upon observed periodic variability, this periodicity thus far spans only two cycles, and correlated noise processes intrinsic to AGN can mimic such periodicity for small numbers of observed cycles if the correlation time of the noise is of order the temporal baseline of observations or longer \citep{Vaughan+2016}. In this work we aim simply to test whether or not the observed sinusoidal variations and peaks in the light curve could be caused by a SMBHB Doppler boost plus self-lensing model.

While by eye, the sinusoidal nature of the Ark 120 light curve is suggestive of a putative binary orbit with modest eccentricity -- the first peak being located near the mean of the sinusoidal portion of the light curve -- the location of the second peak, below the mean, motivates us to consider eccentric orbits.

We next considered KIC 11606854 (hereafter "Spikey"), the rare case of a type 1 AGN identified in Kepler's high-quality, high-cadence photometric data. Kepler was launched by NASA to detect Earth-sized and smaller exoplanets in or near habitable zones by searching for transits in stellar light curves. Although only 7 AGN were known to be in Kepler's field of view (FOV) prior to the start of the mission, in recent years, efforts by various groups \citep{carini2012, edelson2012} have led to dozens of AGN being discovered in the Kepler FOV. \footnote{Analysis has also been done on these Kepler AGN in recent years: \cite{kasliwal2015} tests the popular damped random walk (DRW) model of AGN variability \citep{macleod2010, kozlowski2010} and finds that less than half the objects considered are consistent with a DRW, and \cite{smith2018} offers a comprehensive analysis of 21 light curves, power spectral density functions (PSDs), and flux histograms, examining the data for correlations with various physical parameters.} Spikey was identified in \cite{smith2018} by its striking symmetric flare in the center of a rising continuum (Figure \ref{fig:Spikeylc}). While Spikey does not exhibit a periodic light curve, by eye, the non-sinusoidal shape of the light curve and location of the flare imply a high-eccentricity orbit within the Doppler + self-lensing model, which we test below. Spikey is a high-z source, at z = $0.918$.

 %%%%%%%%%%%%%%%%%%%%%%%%%%%%%%%%%%%%%%%%%%%%%%%%
%%% FIGURE: Orbital parameters %%%
%%%%%%%%%%%%%%%%%%%%%%%%%%%%%%%%%%%%%%%%%%%%%%%%
\begin{figure}
\begin{center}$
\begin{array}{c}
\includegraphics[scale=0.40]{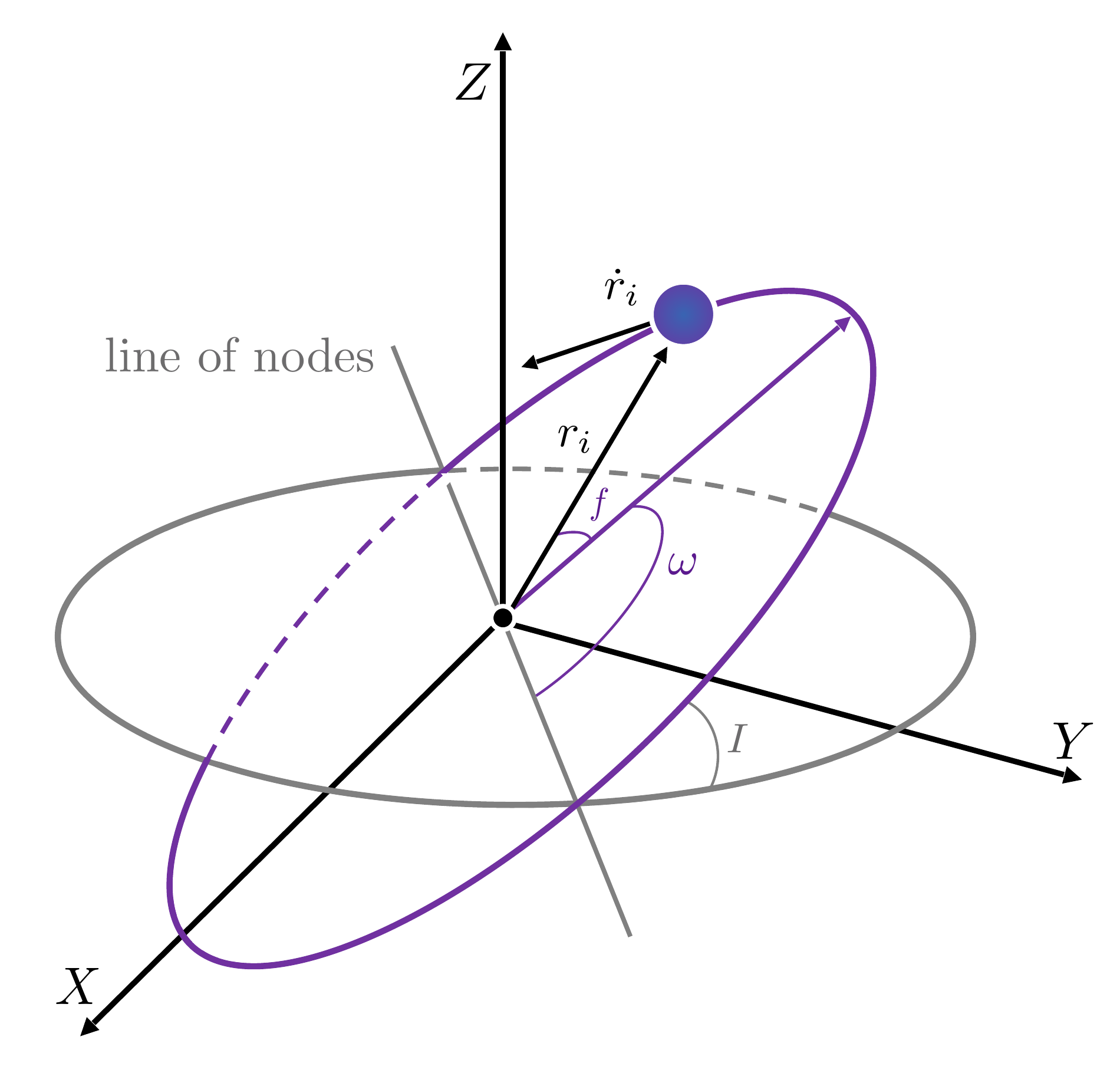} \hspace{20pt} 
\end{array}$
\vspace{-10pt}
\end{center}
\caption{\label{fig:Kepler_diagram}}
Orbital parameters as described in section \ref{ssec:modelcomp}, with the $Z$-axis oriented towards the observer. $\mathbf{r_i}$ is the vector from the center of mass to the i-th mass, $\mathbf{\dot{r}_i}$ the velocity vector, $\omega$ the argument of periapse, $f$ the true anomaly, and $I$ the inclination.
\end{figure}
%%%%%%%%%%%%%%%%%%%%%%%%%%%%%%%%%%%%%%%%%%%%%%%%

\section{Model Description}
\label{sec:model}
\subsection{\label{ssec:modelcomp}Model Components}
In addition to a steady mean flux $F_{\text{0}}$, we consider four components in our light curve model: the Doppler boost, gravitational microlensing, stochastic noise described by the damped random walk (DRW) model, and photometric noise,
\begin{equation} \label{eq:1}
\begin{split}
F_{\text{obs}}&=F_{\text{0}}\times(1+\Delta F_{\text{Doppler}})  \times(1+\Delta F_{\text{lensing}})  \\
& \times(1+\Delta F_{\text{DRW}}) +\Delta F_{\text{photometric}}.
\end{split}
\end{equation} 
For small $\Delta$, we can ignore second and third-order terms between the DRW noise and relativistic effects, giving,
\begin{equation} \label{eq:2}
\begin{split}
\ F_{\text{obs}}&=F_{\text{0}}\times(1+\Delta F_{\text{Doppler}})  \times(1+\Delta F_{\text{lensing}}) \\
&+\Delta F_{\text{DRW}}+\Delta F_{\text{photometric}}.
\end{split}
\end{equation}

A widely used model for stochastic quasar variability in the optical wavelength range is the damped random walk (DRW) model  \citep{kozlowski2010, macleod2010}. DRW variability follows correlated random Gaussian fluctuations that can be described by two scales: $\tau$, the damping or characteristic time scale, and $\sigma$, the long-term standard deviation of variability. The asymptotic value of the structure function SF$_{\infty}$, which measures the mean value of the flux variance for measurements $x(t)$ separated by a given time interval $\Delta t$ \citep{hughes1992}, at large $\Delta t$, is related to the $\sigma$ parameter as SF$_{\infty} = \sqrt[]{2}\sigma$. We adopt $\tau$ and SF$_{\infty}$ as the two main stochastic quasar variability model parameters. The power spectral density (PSD) for the DRW is then given by,
\begin{equation} \label{eq:3}
\ \text{PSD}(f)=\frac{\tau^2 \text{SF}^2_{\infty}}{1+(2\pi f\tau)^2}.
\end{equation}
The DRW has a PSD $\propto f^{-2}$, corresponding to red noise at high frequencies ($f > (2\pi\tau)^{-1}$), flattening to a constant, white noise, at low frequencies ($f < (2\pi\tau)^{-1}$). The break frequency is defined as $(2\pi\tau)^{-1}$, where SF($\Delta t$) flattens to SF$_{\infty}$.

The relativistic Doppler boost as a cause of periodicity in SMBHBs is discussed in the case of quasar PG 1302-102 in \citet{dorazio2015}, and in the case of the periodic-light-curve SMBHB candidates in \citet{charisi2018}.
\cite{dorazio2015} proposes that because optical and UV emissions likely arise from gas bound to the individual BHs, the luminosity of the brighter mini-disk (typically that of the faster moving secondary SMBH) would be Doppler boosted, with the Doppler-induced variability being dominant to hydrodynamically-introduced fluctuations in unequal-mass binaries \citep{D'Orazio:CBDTrans:2016}. Because the number of photons, proportional to $F_{\nu}/\nu^{3}$ with $F_{\nu}$ the apparent flux at a fixed observed frequency $\nu$, is Lorentz invariant, it follows that $F_{\nu}$ is modified from the flux of a stationary source $F^0_\nu$. Defining $D=[(1-\beta^2)^{-1/2}(1-\beta_{\parallel})]^{-1}$, with $\beta=v/c$, $c$ the speed of light, and $\beta_{\parallel}$ the component along the line of sight, and assuming an intrinsic power-law spectrum $F_{\nu}^{0}\propto\nu^{\alpha}$, the apparent flux at a fixed observed frequency is modified to $F_\nu = D^3 F^0 _{D^{-1}\nu} = D^{3-\alpha}F_0$. To first order in $\beta$, this causes a modulation of $F_{\nu}$ by a fractional amplitude $\Delta F_{\nu}/F_{\nu} = \pm(3-\alpha)(v/c)\,\mathrm{cos}\,\phi \,\mathrm{sin}\,I$, with $I$, $\phi$, and $v$ the orbital inclination ($I=\pi/2$ denoting an edge-on binary), phase, and three-dimensional velocity. .

\cite{murray2010} gives the radial velocity, the projection of the velocity vector onto the line of site, as:
\begin{equation} \label{eq:4}
\ v_{r,i}=\mathbf{\dot{r}_i}\cdot\mathbf{\hat{Z}}=V_{Z}+K_i(\mathrm{cos}(\omega+f)+e\,\mathrm{cos}\,\omega),
\end{equation}
where the subscript $i$ denotes the $i^{\rm{th}}$ binary component, 
$\mathbf{r_1}$ denotes the vector from the center of mass to the more massive primary,
the $Z$-axis is oriented toward the observer, $V_{Z}=\mathbf{V}\cdot\mathbf{\hat{Z}}$ is the proper motion of the barycenter, $\omega$ is the argument of periapse, and $f$ is the true anomaly, as shown in Figure \ref{fig:Kepler_diagram}. 
For the secondary,
\begin{equation} \label{eq:5}
\ K_2=\frac{2\pi}{T}\frac{M_1}{M_1+M_2}\frac{a\,\mathrm{sin}\,I}{\sqrt{1-e^2}},
\end{equation}
with $T$ the period of the orbit, $a$ the semi-major axis of the elliptical orbit, and $e$ the eccentricity. We solve for the radial velocity for both binary components, $v_{r1}$ and $v_{r2}$, using $K_1 = qK_2$ for binary mass ratio $q\equiv M_2/M_1 \leq 1$. We introduce an additional variable 
$f_L\equiv L_2/(L_1 + L_2)$, 
the luminosity ratio, and assume that emission from both black holes share the same spectral index $\alpha$.

\cite{gaudi2010} reviews the fundamental concepts of microlensing. 
The characteristic scale of gravitational lensing is the angular Einstein radius,
\begin{equation} \label{eq:7}
\ \theta_{E}\equiv\bigg(\frac{4GM}{D_{\text{rel}}c^2}\bigg)^{1/2},
\end{equation}
where $D_{\text{rel}}^{-1}\equiv D_{l}^{-1}-D_{s}^{-1}$, $D_{l}$ and $D_{s}$ are the distances to the lens and source, respectively, and $M$ is the mass of the lens. For binary self lensing, $D_{\text{rel}}$ is simply the time-dependent distance between the two binary components along the line of sight (separation in the Z-direction) divided by the squared distance to the binary. A strong lensing event occurs when the source is within one Einstein radius of the lens. The width of the lensing flare depends on the orbital separation relative to the Einstein radius. 
Defining $u=\delta/\theta_{E}$, for angular separation $\delta$ between lens and source,
we may write the magnification due to gravitational lensing as,
\begin{equation} \label{eq:8}
\ A(u)=\frac{u^2+2}{u\sqrt{u^2+4}}.
\end{equation}

To calculate the flux from what we refer to as our Doppler + self-lensing model, we write,
\begin{equation} \label{eq:6}
(1+\Delta_{\text{Doppler}})  (1+\Delta_{\text{lensing}}) = (1-f_L) D^{3-\alpha}_{1} A_1 + f_L D^{3-\alpha}_{2} A_2.
\end{equation} 
where $D_1$ ($D_2$) is the Doppler factor due to the line of sight velocity of the primary (secondary), and $A_1$ ($A_2$) is the lensing magnification factor when the secondary (primary) acts as the lens. The Doppler + self-lensing flare model has 10 parameters, listed in Table \ref{tab:parameters}.

We use a Python implementation of Goodman \& Weare's Affine Invariant Markov chain Monte Carlo (MCMC) Ensemble sampler \citep{goodman2010}, emcee \citep{foremanmackey2013}, to fit our Doppler + self-lensing model to the light curves. We maximize the likelihood function, 
\begin{equation} \label{eq:9}
\ \text{ln }p(y|\{x_i\})=-\frac{1}{2}\sum_n\bigg[\frac{(y_n-m_n)^2}{\sigma_n^2}+\text{ln }\sigma_n^2\bigg],
\end{equation}
where $\{x_i\}$ is our set of 10 variables, $y_n$ is the light curve data, $m_n$ is the model given $\{x_i\}$, and $\sigma_n$ is the photometric error. For Ark 120 we use published errors on each data point for $\sigma_n$. For Spikey, we bin the light curve data in bins of $\sim0.5$ days during the spike and bins of $\sim9$ days around the spike, and use the standard deviation of each bin for the error $\sigma_n$. This is done to favor models that fit the flare as opposed to just the Doppler part of the light curve. We assume the noise on the Kepler photometric data to be Gaussian. For both Ark 120 and Spikey, the light curves are fit by the Doppler + self-lensing model using emcee to find the 10 parameters and their uncertainties listed in Table \ref{tab:parameters}.

For the more promising Doppler + self-lensing candidate, Spikey,
we again use emcee to fit the DRW model (Eq. \ref{eq:3}) to the light curve to recover an additional two parameters, $\tau$ and SF$_{\infty}$, from which we can also calculate $\sigma=\text{SF}_{\infty}/\sqrt{2}$. For modeling a DRW process we follow \citep{kozlowski2010} and use the likelihood function,
\begin{equation}
 \mathcal{L}_{\rm DRW} = \left| \textrm{Cov} \right|^{-1/2} \left| L^T \textrm{Cov}^{-1} L \right|^{-1/2}\exp\left( -\frac{ \boldsymbol{Y^{T}}(\textrm{Cov})^{-1}\boldsymbol{Y}}{2}\right).
\end{equation}
Here $L$ is a vector of ones with length equal to the number of data points. $\textrm{Cov}$ is the time-domain covariance matrix comprised of DRW and photometric noise contributions, given by,
\begin{equation} \label{eq:10}
\ \textrm{Cov}_{ij}=\sigma^{2}\textrm{exp}\bigg[\frac{-|t_i-t_j|}{(1+z)\tau}\bigg] + \sigma^2_n \delta_{ij}, \
\end{equation}
where $\delta_{ij}$ is the Kronecker-Delta. 
The vector of residuals, $\boldsymbol{Y} = \boldsymbol{O} - \boldsymbol{M}$ , is constructed from the observed flux $\boldsymbol{O}$ and flux predicted in the model $\boldsymbol{M}$.

We perform basic model selection by applying the DRW model to two different versions
of the light curve: 1) the un-adjusted light curve and 2) the model subtracted light curve.
In the second case, the maximum-likelihood
Doppler + self-lensing model is subtracted before sampling the posterior. 
To assess which of the models is favored by the data, we calculate the Bayesian Information Criterion (BIC) for each model and compare, 
\begin{equation}
\rm{BIC} = k\ln(N) - 2 \ln(\mathcal{L}_{\rm max}),
\label{eq:BIC}
\end{equation}
where $k$ is the number of model parameters (12 for Doppler + self-lensing + DRW, 2 for DRW only), $N$ is the number of data points, and $\mathcal{L}_{\rm max}$ is the maximum likelihood for a given model. A model with a lower BIC is favored, with BIC differences of $\gtrsim6$ being significant.

\begin{table*}
\begin{tabular}{l|l|l|l|l|l}
            &         &  \textbf{Ark 120}    &            &     \textbf{Spikey}  &  \\
  Parameter & Meaning & Prior Range & Parameters & Prior Range & Parameters  \\
\hline
\hline
  $v_z$ [c] & velocity of barycenter along line of sight & [-1, 1] & $0.074^{+0.150}_{-0.063}$ & [-1, 1] & $0.000^{+0.002}_{-0.003}$ \\[0.1 cm]
  $\omega$ [rad] & argument of periapse & [0, $2\pi$] & $4.776^{+1.514}_{-1.144}$ & [0, $2\pi$] & $1.477^{+0.088}_{-0.081}$  \\[0.1 cm]  
  $e$ & eccentricity & [0, 1] & $0.081^{+0.098}_{-0.058}$ & [0, 1] & $0.524^{+0.042}_{-0.043}$ \\[0.1 cm]
  $T$ [yrs] & period & [18, 25] & $19.527^{+0.635}_{-0.594}$ & [0, 3] & $1.144^{+0.031}_{-0.029}$ \\[0.1 cm]
  $\cos{I}$ [rad] & inclination & [-1, 1] & $ 0.281^{+0.482}_{-0.997}$ & [-1, 1] & $0.140^{+0.027}_{-0.022}$ \\[0.1 cm] 
 $\log{\left(M_1/\Msun\right)}$ & mass of primary BH & [5,11] & $9.859^{+0.714}_{-0.728}$ & [5,11] & $7.4^{+0.2}_{-0.2}$ \\[0.1 cm]
  $\log{\left(M_2/\Msun\right)}$ & mass of secondary BH & [5,11] & $6.778^{+2.024}_{-1.278}$ & [5,11] & $6.7^{+0.5}_{-0.7}$ \\[0.1 cm]
  $f_L$ & luminosity ratio & [0,1] & $0.654^{+0.239}_{-0.287}$ & [0,1] & $0.89^{+0.08}_{-0.14}$  \\[0.1 cm]
  $t_0$ [yrs] & arbitrary reference time & [-10, 30] & $23.167^{+4.536}_{-3.559}$ & [-3, 3]& $1.693^{+0.032}_{-0.029}$ \\[0.1 cm]
  $\alpha$ & spectral index & [-6, 6] & $-1.555^{+2.180}_{-2.729}$ &[-4, 4] & $2.09^{+0.18}_{-0.29}$ \\[0.1 cm]
\end{tabular}
\caption{Recovered model parameters, uncertainties and assumed priors. All values are in the rest frame of the source. Here we quote parameter values derived from the $50\%$, $16\%$, and $84\%$ quantile estimates, see the Appendix. The results for Spikey (Ark 120) utilize the last 20,000 (2,000) steps of the MCMC walker chains.
}
\label{tab:parameters}
\end{table*}

\subsection{\label{ssec:modellc}Model Light Curves}

D18 describe self-lensing as a unique signature of accreting SMBHBs. They show
that self-lensing is expected in a few to tens of percent of accreting SMBHBs,
depending on binary parameters and assuming a circular orbit throughout.  They
compare lensing flares for different mass ratios  and find that for extreme
mass ratio cases $q \lsim 0.05$, a significant lensing flare occurs only as
the secondary passes behind the more massive primary. For larger mass ratios,
a second lensing flare can occur as the primary passes behind the secondary,
assuming the primary is accreting as well.

\begin{figure*}
\begin{center}$
\begin{array}{c c}
\includegraphics[scale=0.28]{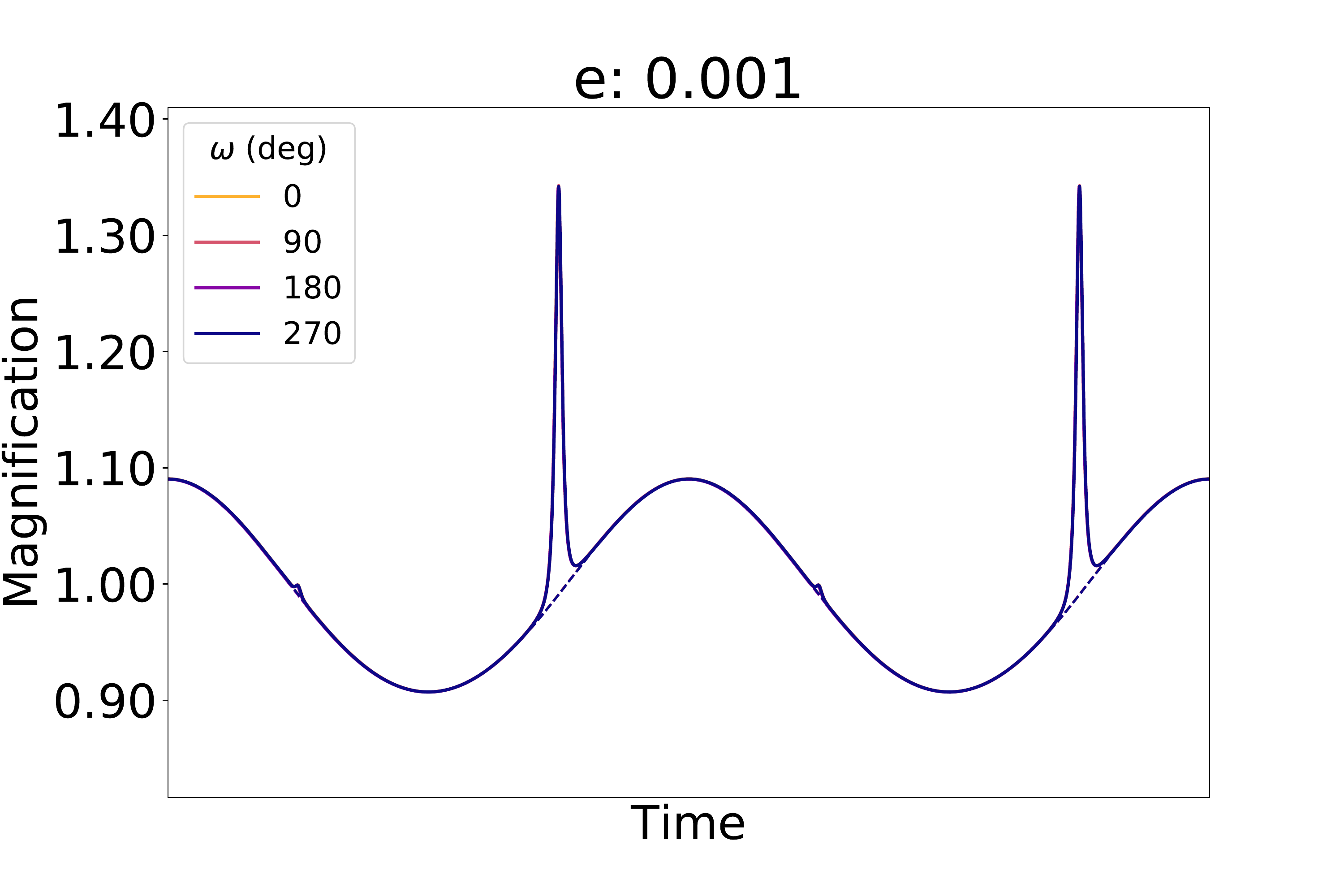} & \includegraphics[scale=0.28]{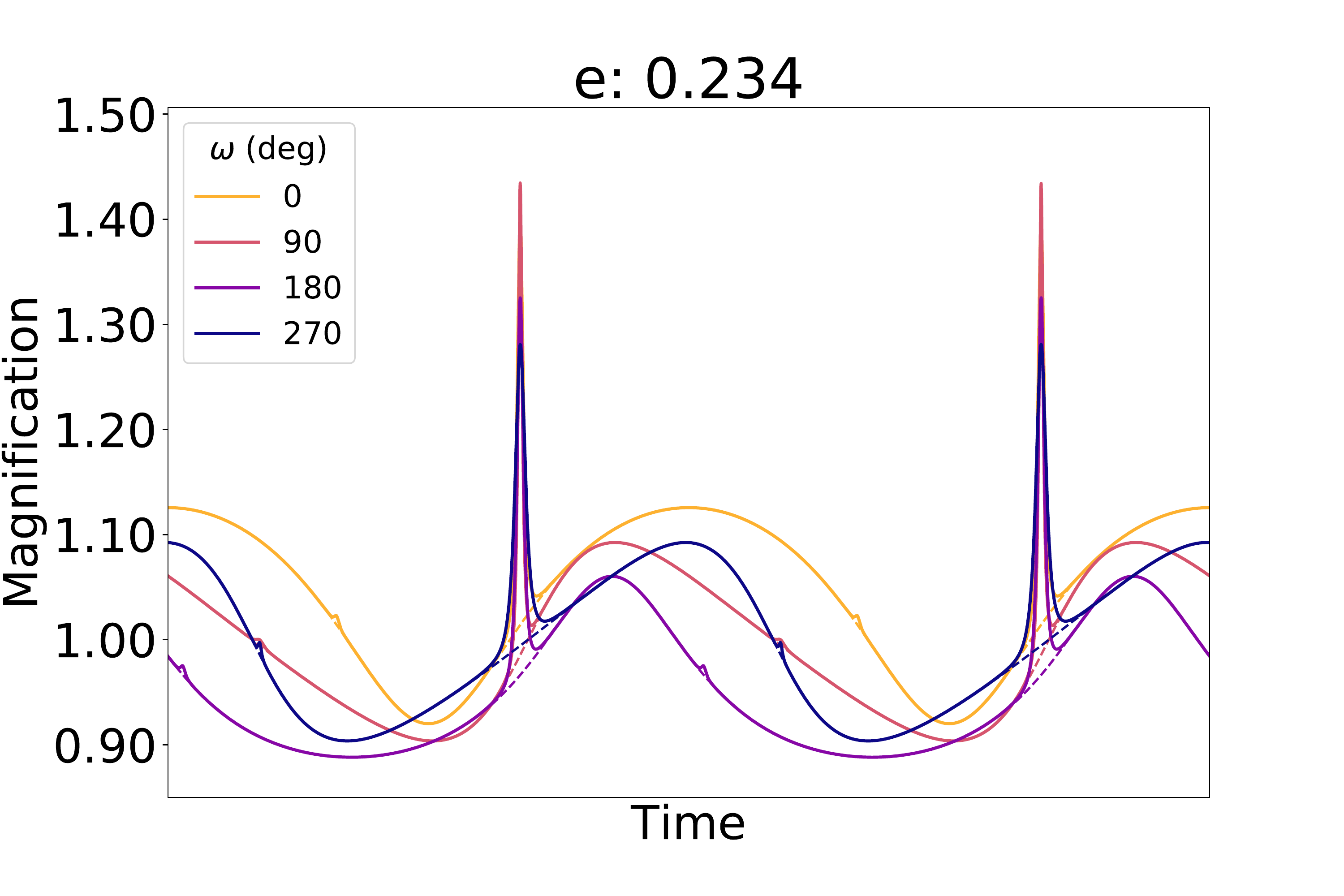}  \\ 
\hspace{7pt}
\includegraphics[scale=0.28]{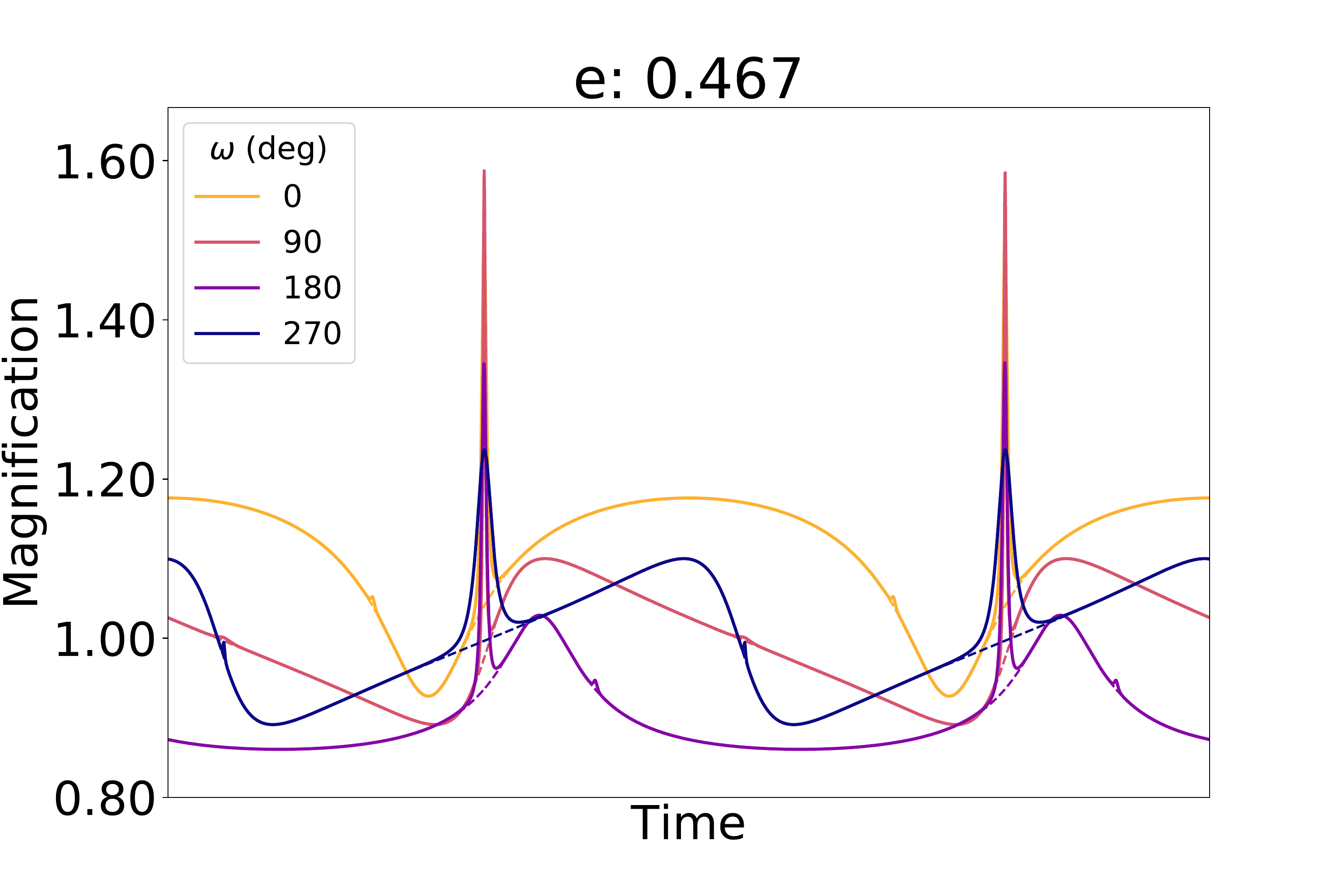} & \includegraphics[scale=0.28]{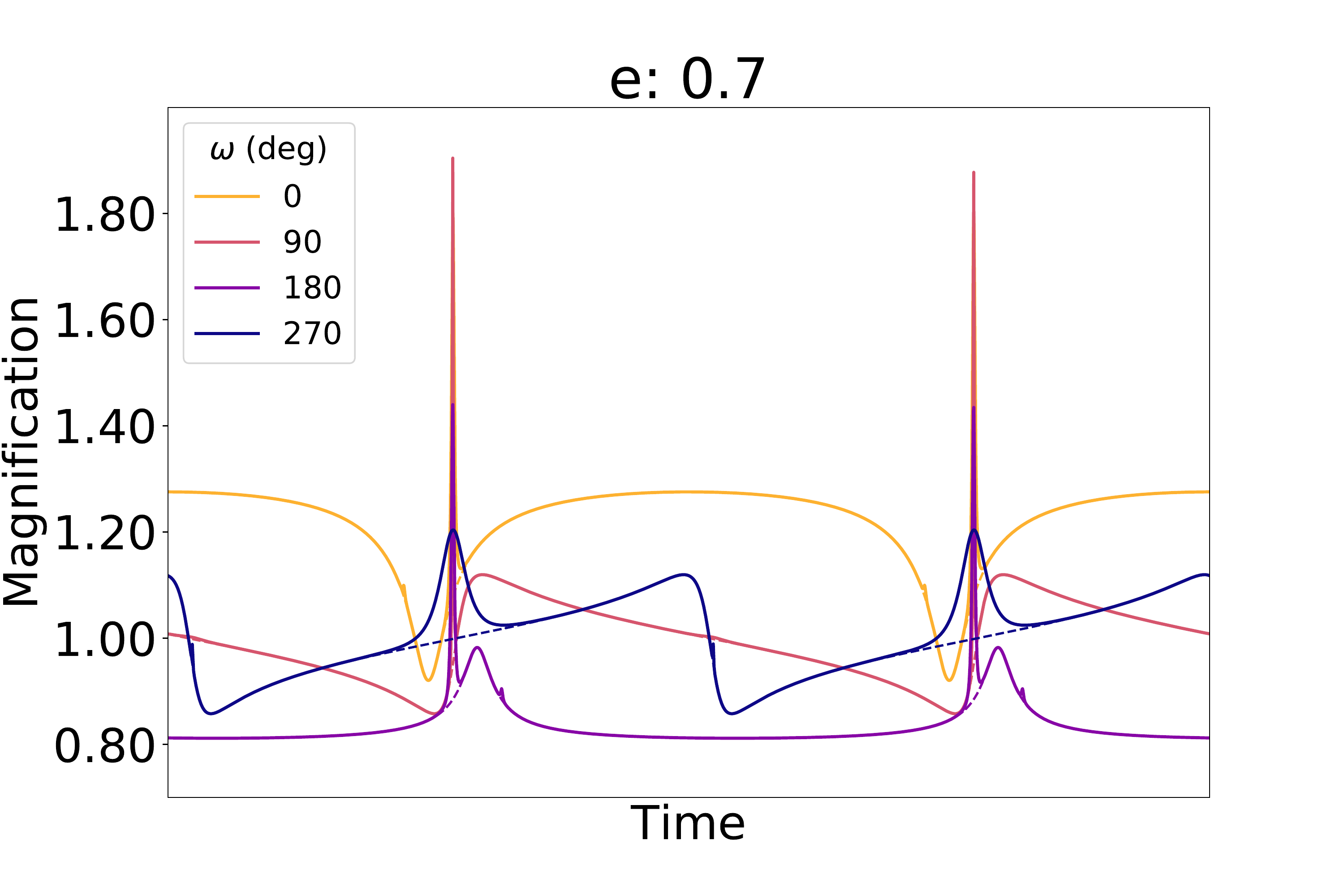}  \\ 
\end{array}$
\end{center}
\vspace{-15pt}
\caption{
Example light curves for $q\equiv M_2/M_1 = 0.1$, $f_L=0.7$, $I=87^{\circ}$, fixed eccentricity $e$ in each subplot, and varying argument of periapse $\omega$. Parameters are chosen to allow for a second lensing flare to appear. Dotted lines follow the light curve if there were no lensing flares. Light curve traces are shifted along the $x$-axis to align the primary lensing flares in each plot.  }
\label{fig:example_lcs}
\end{figure*}

Motivated by the light curves of the two systems studied in this work, we
extend  the lensing model for a wider range of situations. In particular, we
focus here on the dependence on two new parameters, which have not been
previously discussed for Doppler + self-lensing models: the eccentricity $e$
and the argument of periapse $\omega$. In Figure \ref{fig:example_lcs}, we
show light curves for four example binaries with inclination $I=87\degree$
with varying $e$ and $\omega$, for $M_1=5\times10^8\,\Msun{}$,
$M_2=5\times10^7\,\Msun{}$ ($q=0.1$), $T\approx20$ years, and $f_{L}=0.7$,
chosen to allow a second lensing flare to appear (assuming that both black
holes are accreting). In plotting the light curves, we shift the light curves
along the $x$-axis to align the primary lensing flares (where we refer to the
primary lensing flare as the highest magnification flare generated when the
primary acts as the lens). For circular orbits, the line-of-sight
conjunctions, which correspond to the peaks of the lensing spikes, always
coincide with the average flux with no Doppler boost because of the vanishing
line-of-sight velocity. We show in Figure \ref{fig:example_lcs} that for
$e>0$, this is only true if $\omega=90\degree\text{ or }270\degree$, for which
the binary reaches periapse at its northmost or southmost distance from the
plane of reference, respectively. The lensing spike is offset from the average
flux for all other values of $\omega$. When $\omega=0\degree\text{ or
}180\degree$, the binary reaches periapse when it is crossing the plane of
reference from South to North or North to South, respectively.

For highly eccentric orbits, the shape of the light curve and flare
magnification can vary dramatically with the argument of periapse $\omega$.
This is because the Einstein radius (Eq. \ref{eq:7}), which sets the flare
magnification and width, depends on the line of sight projected distance
between the lens and source through the binary orbit, which in turn changes
with both $e$ and $\omega$.

For eccentric orbits, the orbital velocity of both binary components is faster
at periapse and slower at apoapse than for a circular orbit with the same
period. Hence the widest flares occur during apoapse conjunctions and the
narrowest occur at periapse conjunctions. At $\omega=270\degree$, the
secondary is lensed at apoapse and the primary is lensed at periapse,
resulting in the widest primary flares and narrowest secondary flares.
In contrast, at $\omega=90\degree$, the secondary is lensed at periapse and the
primary is lensed at apoapse, resulting in the narrowest primary flares and
widest secondary flares. For $\omega$ near $0\degree$ or $180\degree$, both
lensing flares occur between the apoapse and periapse, but closer to the
periapse, resulting in relatively narrow flares.

The binary components in an eccentric orbit have larger (smaller) separation
at apoapse (periapse) than a binary with the same orbital period on a circular
orbit. This has two consequences for the magnification. On one hand, lensing
events peaking at periapse will have $\propto \sqrt{(1+e)/(1-e)}$ times
smaller Einstein radii than those peaking at apoapse. On the other hand, the
angular separation of source and lens is a factor of $\propto (1+e)/(1-e)$
larger at apoapse than it is at periapse. Hence, the strongest lensing flares
occur when the secondary passes behind the primary at periapse; the weakest
when the primary passes behind the secondary at apoapse.

This combined behaviour can be seen in the top right and bottom panels of
Figure \ref{fig:example_lcs}. The $\omega=90^{\circ}$ curve denotes orbits
where the secondary passes behind the primary at periapse, yielding the
highest-magnification, narrowest primary lensing flares. During the same
orbit, the primary is lensed by the secondary at apoapse, yielding the widest,
lowest magnification secondary lensing flares. The opposite case yielding the
widest, lowest magnification primary flares and the narrowest highest
magnification secondary flares is seen for the $\omega=270^{\circ}$ case.

As is well known, the lens mass and binary inclination can also change the
width and magnification of the lensing flare. Hence, we expect the
eccentricity to add further degeneracy in fitting for these parameters in the
next section.

%%%%%%%%%%%%%%%%%%%%%%%%%%%%%%%%%%%%%%%%%%%%%%%%
%%% FIGURE: Example light curves%%%
%%%%%%%%%%%%%%%%%%%%%%%%%%%%%%%%%%%%%%%%%%%%%%%%

\section{\label{sec:results}Results}
\subsection{\label{ssec:ark120results}Ark 120}

\begin{figure}
\begin{center}$
\begin{array}{c}
\includegraphics[scale=0.28]{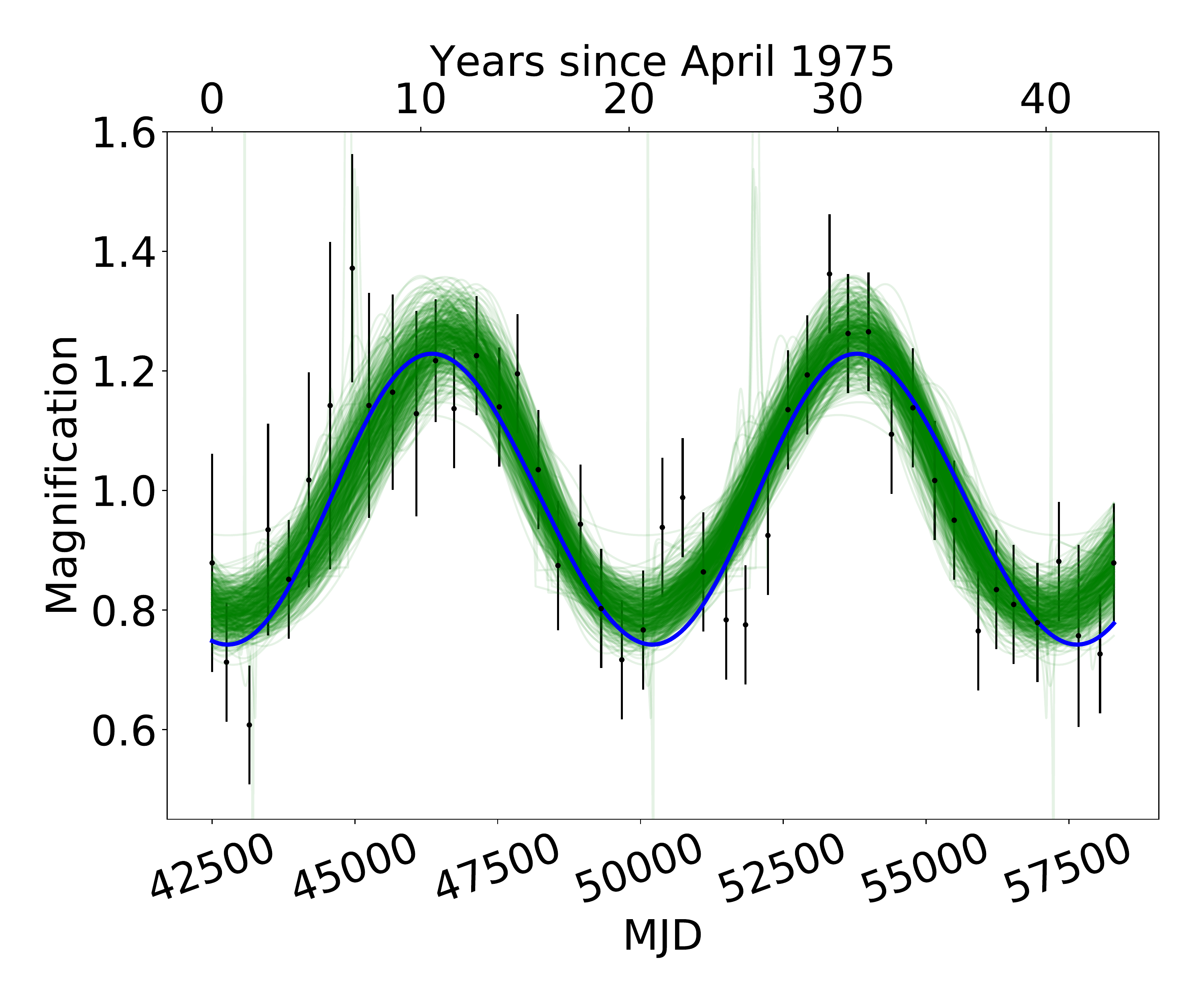} \hspace{20pt} 
\end{array}$
\vspace{-15pt}
\end{center}
\caption{\label{fig:Ark120lc}
Optical light curve for Ark 120, overlaid with the maximum-likelihood Doppler + self-lensing model light curve in blue, and model realizations for 0.95 * (number of walkers) sets of parameters randomly drawn from the set of samples to represent the 95\% uncertainty, in green. For this fit, emcee was run for 20,000 steps with 500 walkers.
  }
\end{figure}

Figure \ref{fig:Ark120lc} shows the optical light curve for Ark 120, overlaid
with the maximum-likelihood Doppler + self-lensing model light curve in blue.
Table \ref{tab:parameters} lists the $50\%$, $16\%$, and $84\%$ quantile
parameter values recovered by the MCMC posterior sampling. In green, we plot
model realizations for 0.95 * (number of walkers) sets of parameters randomly
drawn from the emcee samples to represent the uncertainty in the model.

For this fit, emcee was run for 20,000 steps with 500 walkers. Analysis is
carried out on the final $10\%$ of the chains, well after convergence.  The
Doppler-boost model requires a negative (or small positive) $\alpha$ due to
the long period and large amplitude. Publicly available optical spectra,
covering wavelengths around the $H\beta$ line, not far from the $V$ band,
exist from 1976 to 2017 \citep{capriotti1982, korista1992, stanic2000,
peterson1998, doroshenko2008}. We use four spectra from \cite{korista1992},
dating from 1981 through 1984, and find that the power-law component of the
continuum is best fit with an average spectral index of $\alpha$ = $2.36$. For
this larger measured value of $\alpha$, a larger binary mass and smaller mass
ratio at fixed orbital period could explain the large amplitude of
Doppler-boost variations. Hence, we repeat our analysis, but with
$\alpha$ fixed to be 2.36. We find  $\log\left[M_1/\Msun \right]\approx
6.5^{+1.4}_{-0.9}$ and $\log\left[M_1/\Msun \right]\approx 10.5^{+0.4}_{-0.7}$,
with the other parameters largely unchanged. This suggest a primary mass
approaching the limit of expected and known supermassive black hole masses,
but not so large that the Doppler model is ruled out. While not theoretically
impossible, this large mass is at odds with a measured value of the central
compact mass within Ark 120. Broad-line-width measurements provide an estimate
of $M_{\text{BH}} = (2.6\pm0.2)\times10^8 \Msun$ \citep{li2017}.
However, systematic uncertainties in central mass estimation from
broad-line measurements, as well as the possibility of a time variable
spectral slope, could alleviate the above consistencies.  Furthermore, we
note that the Kepler bandpass ranges from 420 to 900 nm, while the spectra
from \cite{korista1992} only covered 450 to 550 nm.

Hence, based on the mass inconsistency and in light of these caveats, we disfavor but do not rule out the Doppler hypothesis for Ark 120. Further measurements of the spectral slope and central mass in Ark 120 as well as further optical monitoring can elucidate these issues in the future.

Even in the case that the Doppler-boost scenario is viable, we find that model realizations exhibiting a significant lensing flare are not favored. This can be seen more quantitatively by comparing the BIC (Eq. \ref{eq:BIC}) computed for the Doppler + self-lensing model to the BIC computed for the Doppler-only model. The difference in BIC between the two is consistent with zero, as is expected since both models have the same number of parameters, and find a similar best-fit.

Outside of the Doppler + self-lensing, scenario, we note that periodicity due
to time variable accretion coupled with lensing is not ruled out, but is more
difficult to test as there is no predicted correlation between the sinusoidal
variability and the lens flare; the lens flare could occur at any phase and
amplitude relative to the sinusoidal variability. Regardless, observed
periodically recurring flares in Ark 120 would warrant further investigation
into this possibility.

%%%%%%%%%%%%%%%%%%%%%%%%%%%%%%%%%%%%%%%%%%%%%%%%
%%% FIGURE: Ark 120 %%%
%%%%%%%%%%%%%%%%%%%%%%%%%%%%%%%%%%%%%%%%%%%%%%%%
%%%%%%%%%%%%%%%%%%%%%%%%%%%%%%%%%%%%%%%%%%%%%%%%

\subsection{\label{ssec:spikeyresults}Spikey}
\begin{figure*}
\begin{center}$
\begin{array}{c c}
\includegraphics[scale=0.26]{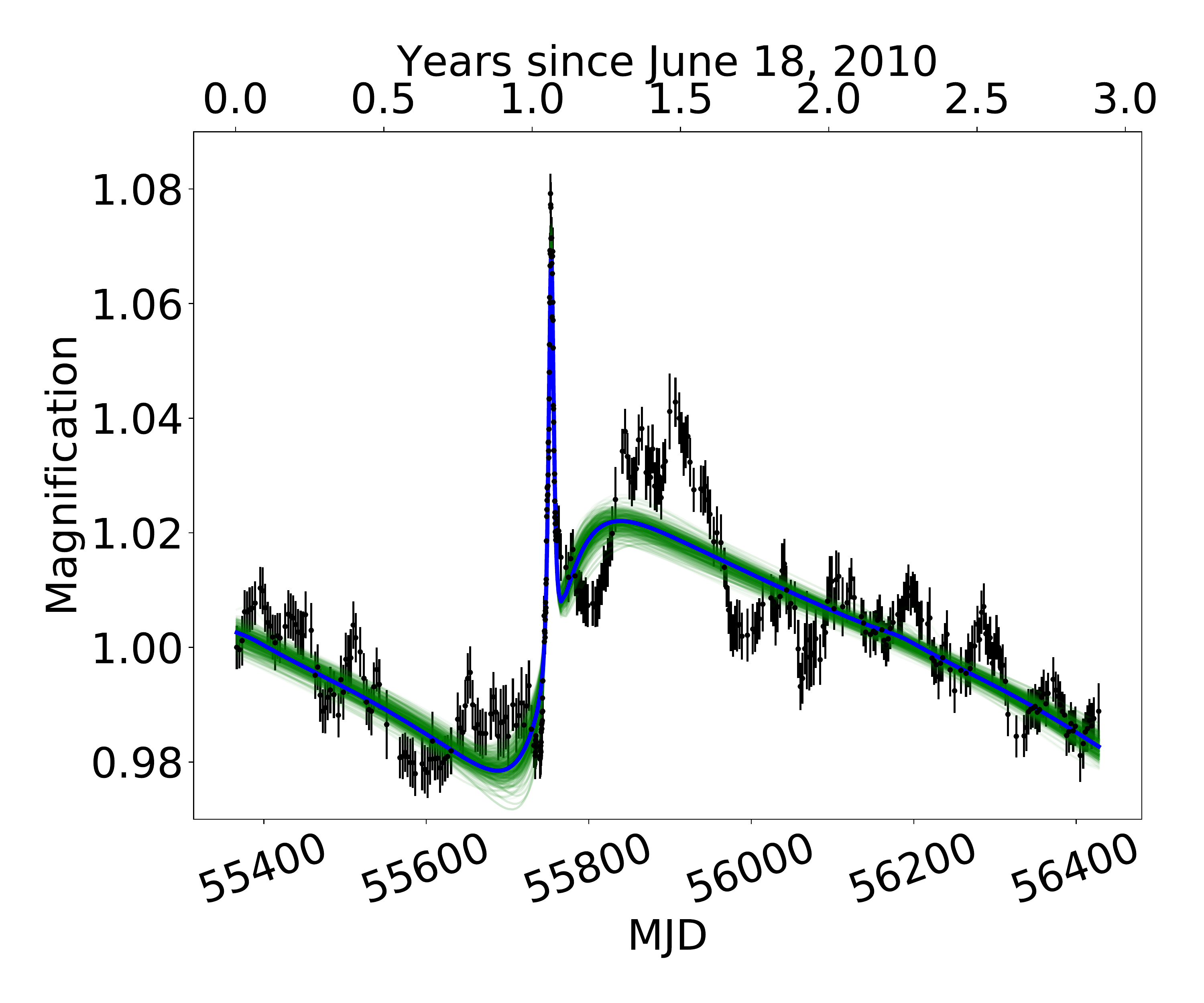} & \includegraphics[scale=0.26]{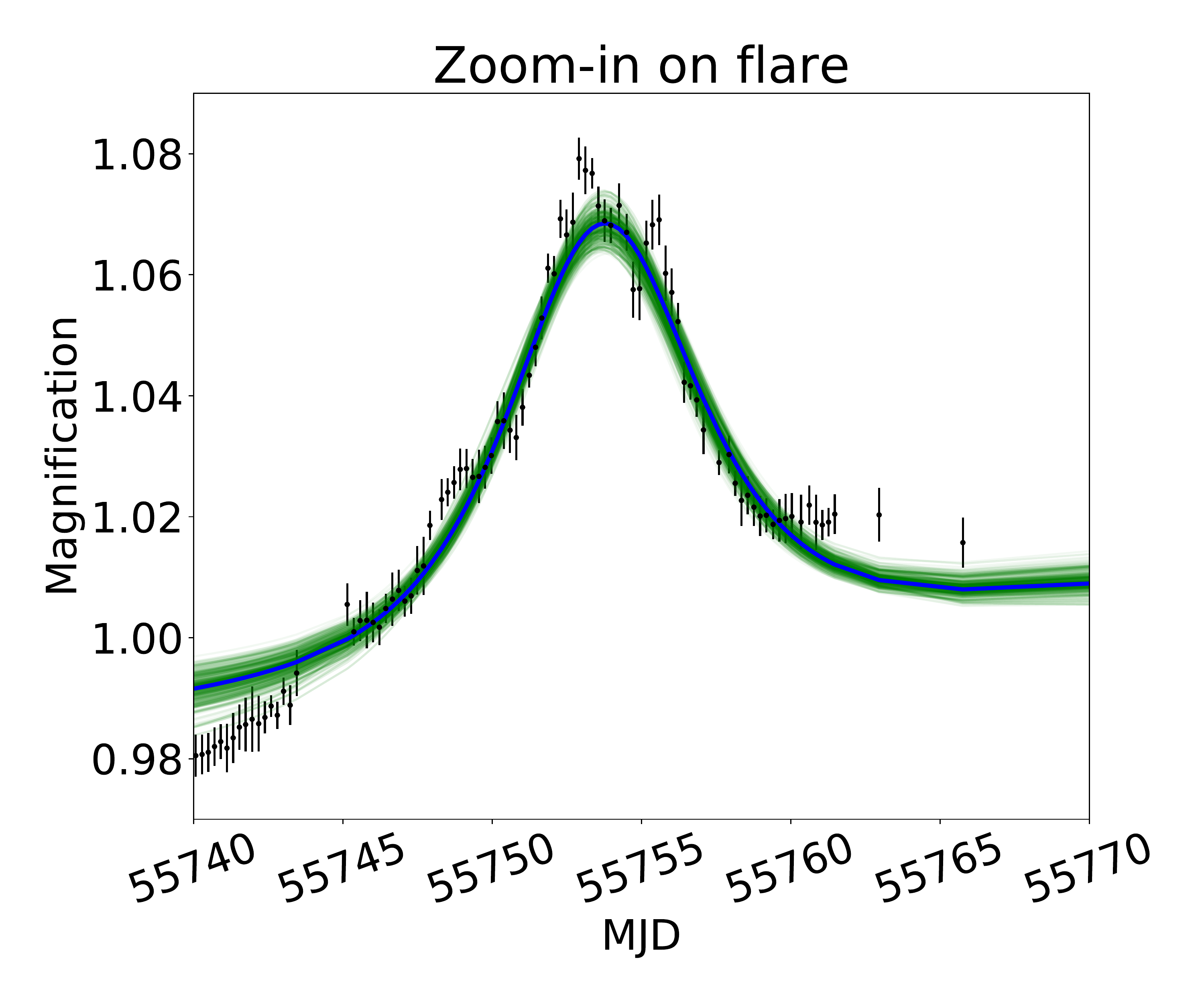}
\end{array}$
\end{center}
\vspace{-15pt}
\caption{
Optical light curve for Spikey, overlaid with the maximum-likelihood Doppler + self-lensing model, in blue, and 95\% uncertainty, in green. Here, emcee was run for 50,000 steps with 500 walkers. On the right we zoom-in on the lensing flare.
\label{fig:Spikeylc}
}
\end{figure*}

Figure \ref{fig:Spikeylc} shows the optical light curve for Spikey overlaid with the maximum-likelihood Doppler + self-lensing model light curve in blue, and Table \ref{tab:parameters} lists the $50\%$ quantile parameters values with errors quoted from the $16\%$ and $84\%$ values. Again, in green we plot model realizations for 0.95 * (number of walkers) sets of parameters randomly drawn from the emcee samples to represent the 95\% uncertainty. emcee was run for 50,000 steps with 500 walkers. The overall shape of Spikey's light curve, without lensing, is non-sinusoidal, suggesting a non-zero eccentricity. The narrowness of the spike, with a width of approximately 10 days, can be due to the Einstein radius being small compared to the orbital separation. In the right panel of Figure \ref{fig:Spikeylc} we see that the symmetric shape of the spike is remarkably close to the symmetric Paczynski-curve shape predicted in the Doppler + self-lensing flare model. 

An  optical  spectrum  was  obtained for Spikey on June 25, 2012, from which a power-law slope is measured to be $\alpha \sim -0.12$.\footnote{The optical spectrum was obtained with the KAST double spectrograph on the Shane 3-m telescope at Lick Observatory. To obtain the spectral index in this region, the two strong emission lines (C~III~$\lambda1909$ and Mg~II~$\lambda2799$) are masked out and the continuum is fit with a linear model using a least-squares method.} This is not consistent with our recovered $\alpha$ parameter for Spikey (Table \ref{tab:parameters}); however, the error on the measured value could be large. Otherwise the discrepancy could arise from a variable spectral slope within the Doppler + self-lensing model.

%%%%%%%%%%%%%%%%%%%%%%%%%%%%%%%%%%%%%%%%%%%%%%%%
%%% FIGURE: Spikey %%%
%%%%%%%%%%%%%%%%%%%%%%%%%%%%%%%%%%%%%%%%%%%%%%%%
%%%%%%%%%%%%%%%%%%%%%%%%%%%%%%%%%%%%%%%%%%%%%%%%

%%%%%%%%%%%%%%%%%%%%%%%%%%%%%%%%%%%%%%%%%%%%%%%%
%%% FIGURE: Spikey prediction %%%
%%%%%%%%%%%%%%%%%%%%%%%%%%%%%%%%%%%%%%%%%%%%%%%%
\begin{figure*}
\begin{center}
\includegraphics[scale=0.42]{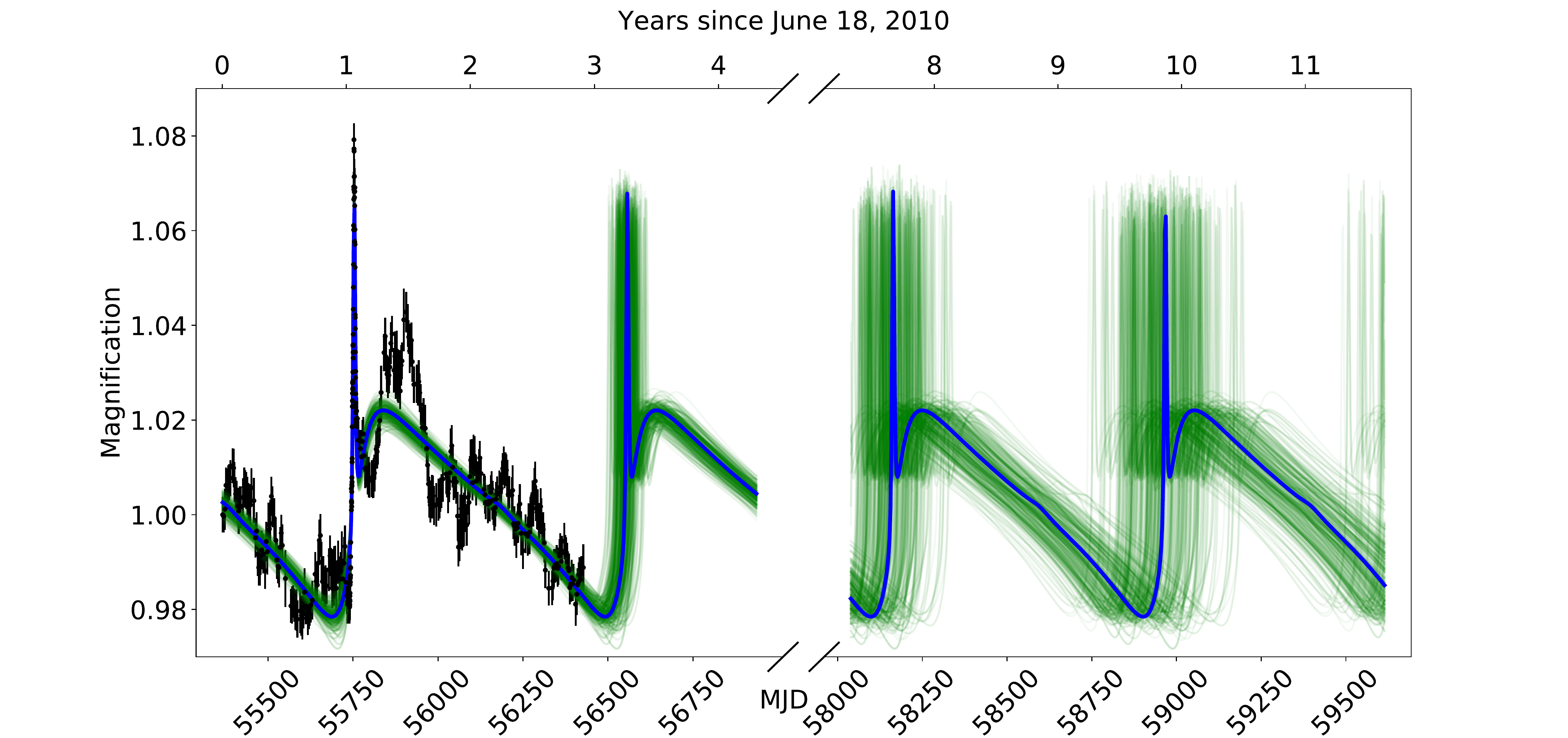}
\end{center}
\vspace{-12pt}
\caption{
Optical light curve for Spikey, with overlaid maximum-likelihood Doppler + self-lensing model and uncertainty extended to show predicted flares in September 2013, February 2018, and April 2020. Cut out of the plot is the predicted flare in December 2015. A July 2022 flare is also predicted.}
\label{fig:Spikeypredict}
\end{figure*}
%%%%%%%%%%%%%%%%%%%%%%%%%%%%%%%%%%%%%%%%%%%%%%%%

To further investigate the plausibility of the Spikey Doppler + self-lensing model, we calculate the BIC (Eq. \ref{eq:BIC}) for DRW models of the light curve with and without the Doppler + self-lensing model (Eq. \ref{eq:4} through \ref{eq:8}) subtracted, as described in \S\ref{ssec:modelcomp}. This allows us to assess whether including the Doppler + self-lensing model improves the fit relative to the DRW model alone.

We find a strong difference in the BIC between the two models of
$\approx 190$, favoring the Doppler + self-lensing + DRW model over the purely
DRW model. The $50\%$ quantile likelihood DRW parameters, with errors from the
$16\%$ and $84\%$ quantile values, for the DRW-only model are $\tau =
586^{+289}_{-337}$ days and $\rm{SF}_{\infty} = 0.052^{+0.012}_{-0.018}$ mag,
and for the Doppler + self-lensing subtracted light curve, $\tau =
434^{+397}_{-365}$ days and $\rm{SF}_{\infty} = 0.034^{+0.013}_{-0.020}$ mag. 
We note that the maximum likelihood parameters are slightly different due to
the non-Gaussian nature of the posterior. For the Doppler + self-lensing
subtracted light curve, a peak in the posterior probability is found at
smaller values of $\tau=31$ days and $\rm{SF}_{\infty}=0.01$ mag. For the
un-subtracted light curve, no such peak is found and the maximum-likelihood
parameters are $\tau=123$ days and $\rm{SF}_{\infty}=0.024$ mag.  This test
confirms that it is not likely for the DRW process to generate such a sharp
feature on short timescales. The priors are chosen so that $\tau$ encompasses
the temporal baseline of the data $\tau=[0,10^3]$ days while
$\rm{SF}_{\infty}$ is essentially unbounded $\rm{SF}_{\infty}=[0,200]$ mag.

\begin{figure}
\begin{center}$
\begin{array}{c}
\includegraphics[scale=0.40]{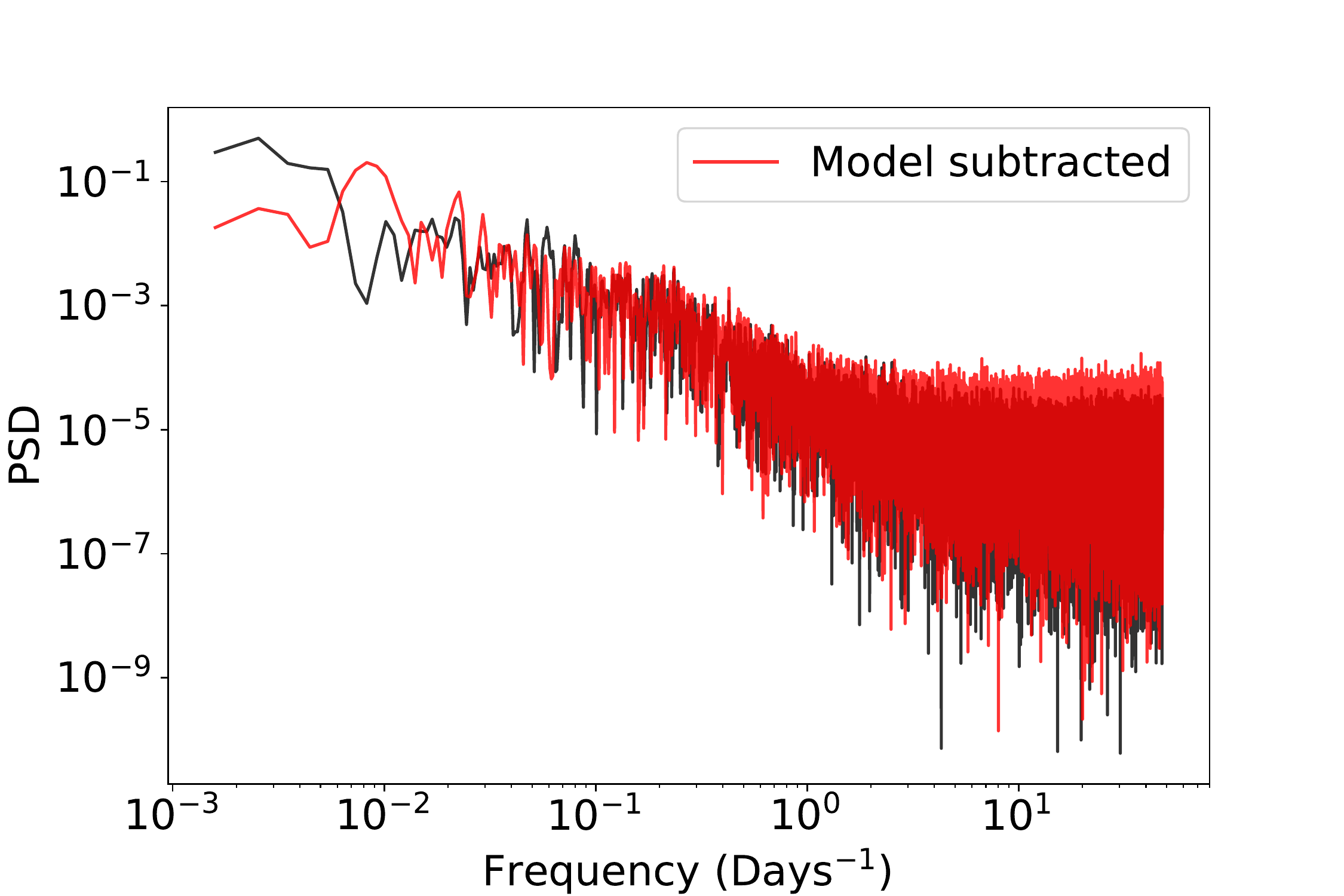} \hspace{20pt} 
\end{array}$
\vspace{-10pt}
\end{center}
\caption{\label{fig:PSD_modelsubtract}
PSD$(f)$ for Spikey data with (red) and without (black) the Doppler + self-lensing model subtracted. With the model subtracted, the PSD$(f)$ begins to show a break, while without the model subtracted, no break was found. 
  }
\end{figure}

In Figure \ref{fig:PSD_modelsubtract}, we plot the PSD$(f)$
for the data with and without the Doppler + self-lensing model subtracted.
With the Doppler + self-lensing model subtracted, the PSD$(f)$ begins to show
a break, as described in \S\ref{ssec:modelcomp}, while previously, no break
was found \citep{smith2018}. The break appears presumably because of the
additional power at the frequency of $f\sim1/\text{yr}$ contributed by the
Doppler modulation.

The most obvious way to test the self-lensing SMBHB hypothesis for Spikey is
to look for periodically recurring lensing flares. In Figure
\ref{fig:Spikeypredict}, we extend the Doppler + self-lensing model for
Spikey's light curve to show future lensing flares through 2020. Three
predicted lensing flares could have occurred since Kepler observed Spikey. The
next putative flares are set to occur in April 2020 and July 2022. We discuss
the possibility of detecting these flares in archival and future observations
below.

\section{\label{sec:discussion}Discussion}
We note that our lensing model assumes a point source, while the optical
emitting region is known to have a finite size of a few hundred gravitational
radii \citep{paczynski1977, roedig2014, AL94}. Assuming black-body
emission from a steady-state, optically thick accretion disk, D18 compute when
the wavelength-dependent size of an accretion-disk source becomes of order the
size of the Einstein radius, and hence falls within the finite-sized source
regime. Using Eq. (9) of D18 with the maximum-likelihood Spikey binary parameters,
$10\%$ accretion efficiency at the Eddington limit (consistent with the
luminosity and mass estimate for Spikey), and considering the $420-900$~nm
Kepler bandpass, we find that the ratio of effective disk size to Einstein
radius ranges from $\sim0.3-0.9$. Hence, the putative accretion disk around the
lensed binary component may act like a finite-sized source at the 
long-wavelength end of the Kepler band.

As D18 show, a finite-sized source results in a lower magnification than for a
point-source, and in extreme cases, a wider lensing flare. This could affect
the mass and eccentricity parameter estimation computed here. However, as
evidenced by Figure 5 in D18, we do not expect this effect to be large,
especially in the marginally finite-sized source regime in the Kepler band
into which Spikey falls.

While a useful estimate of the relevant lensing regime, the above is based on
one source model which would itself depend on the inclination of the disk to
the line-of-sight. Whether or not the source must be treated as a finite
source depends on the unknown emission region structure in the lensed accretion
flow. However, we note that the behavior of finite source lensing and the
propensity of accretion flows to be hotter closer to the central compact
object suggest that finding wider, lower magnification symmetric flares at
longer wavelengths is indicative of lensing. If the binary self-lensing
hypothesis could be confirmed for Spikey, then multi-wavelength observations
of a flare would teach us a great deal about the emission region geometry.

Finally, our model for Spikey predicts a relativistic orbital precession of
the argument of periapse by $\sim1.3^{\circ}$ per orbit, which would
alter the timing of the next flare by approximately 1.5 days per orbit. While
this does not greatly affect the prediction for the time of the next flare, it
does present the exciting prospect of tracking general relativistic effects on
the orbit with self-lensing.

\subsection{Other data and first X-ray observations of Spikey}

\begin{figure*}
\begin{center}$
\begin{array}{c}
\includegraphics[scale=0.5]{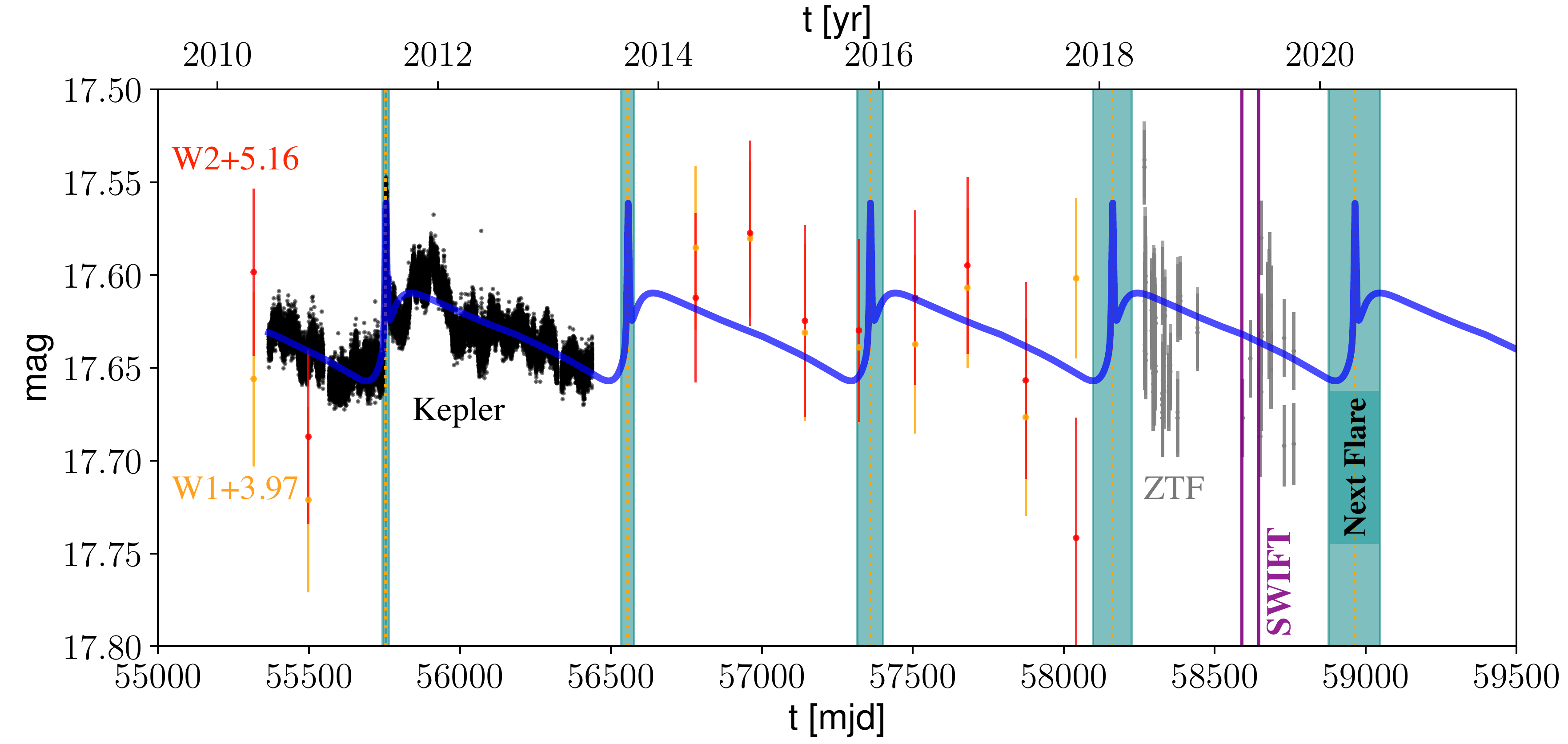} 
\end{array}$
\vspace{-15pt}
\end{center}
\caption{
Archival data on Spikey along with the dates of the first X-ray data obtained with Swift (purple vertical lines). Shaded teal regions are the projected dates of lensing flares, the next (at writing) being in 2020. Not shown are unreleased TESS and Gaia observations as well as proposed Chandra observations during the projected next flare.
  }
\label{fig:Alldat}
\end{figure*}

We have used only Kepler data in vetting the Doppler + self-lensing model for
Spikey. While other data exists, none of it has a high enough cadence or
photometric precision to further constrain or rule out our model. Figure
\ref{fig:Alldat} shows optical data from the Zwicky Transient Facility
\citep[ZTF; which is unfortunately sparse due to positioning of Spikey on a
chip gap,][]{ZTF:2014} and IR data from the Wide-field Infrared Survey Explorer
(WISE). In addition, the TESS satellite has observed Spikey during the months
of June through September 2019 and Gaia has observed Spikey since 2015, and
will continue to do so.

We additionally found data from the ASAS-SN photometric database; however, it
is likely that the PSF for ASAS-SN is too large to isolate emission from
Spikey alone. The data consists largely of upper limits except for a number of
widely varying detections ranging three orders of magnitude in brightness. As
there are multiple stars in this magnitude range near Spikey, and as Spikey is
near the magnitude limit for ASAS-SN ($\sim 18^{\rm{th}}$ magnitude), we do
not include these data.

Future planned observations tailored to observing a repeating flare, as well
as further data from ZTF, TESS, and Gaia, will be vital in ruling out or
confirming the Doppler + self-lensing scenario for Spikey. Gaia epoch data, when it
is released, will likely sample Spikey's light curve with high enough cadence
to rule out or confirm the self-lensing scenario.  With this, as well as the
benefits of multi-wavelength (especially X-ray) observations for confirming
and learning from the self-lensing scenario in mind, we have obtained X-ray
observations of Spikey with Swift and have scheduled observations with Chandra.

Initial X-ray observations of Spikey were performed with Swift during Cycle 15
on 2019 March 16 and 2019 June 22 for 14.4\,ks and 14.0\,ks, respectively. The
observations were intended to measure the X-ray properties of the source and
establish X-ray light curve data. Spikey was confirmed to be a moderately
bright X-ray source with an average count rate of \mbox{0.01\,cts\,s$^{-1}$}
with the Swift X-Ray Telescope (XRT) instrument. XRT spectra were fit with a
{\tt phabs$\cdot$zpowerlaw} model using Cash statistics (cstat) to determine
the spectral slope $\alpha$ and 0.5--7.0\,keV flux $F_{0.5-7.0\rm\,keV}$ in
each observation, where the parameters are defined the same as in Section~3.
The Galactic column density was set to $n_{\rm H} =
6.8\times10^{20}\,\text{cm}{^2}$ as extrapolated from \cite{Dickey1990}. 
Best-fit results for the March 16 observation were \mbox{$\alpha=0.56\pm0.12$} and
\mbox{$F_{0.5-7.0\rm\,keV} =
6.2\pm1.0\times10^{-13}\rm\,erg\,cm^{-2}\,s^{-1}$}, while the June 22 best-fit
results were \mbox{$\alpha=0.67\pm0.18$} and \mbox{$F_{0.5-7.0\rm\,keV} =
4.5\pm0.9\times10^{-13}\rm\,erg\,cm^{-2}\,s^{-1}$}. The flux estimates
indicate a decrease in brightness over time, which is broadly consistent with
the decreasing intensity expected for Spikey during this period.

This first detection of X-rays from Spikey is promising for upcoming Chandra
observations presently scheduled for the predicted flaring period in 2020.
These observations will be sensitive enough to detect a $5\%$ increase in
brightness, and hence detect the putative next flare. Detection of a repeating
flare would provide very strong evidence for the SMBHB and self-lensing
scenario, and importantly, non-detection of a flare will remove evidence for
the SMBHB hypothesis.

%%%%%%%%%%%%%%%%%%%%%%%%%%%%%%%%%%%%%%%%%%%%%%%%
%%% FIGURE: PSD(f) %%%
%%%%%%%%%%%%%%%%%%%%%%%%%%%%%%%%%%%%%%%%%%%%%%%%
%%%%%%%%%%%%%%%%%%%%%%%%%%%%%%%%%%%%%%%%%%%%%%%%

%%%%%%%%%%%%%%%%%%%%%%%%%%%%%%%%%%%%%%%%%%%%%%%%
%%% FIGURE: Extra DATA %%%
%%%%%%%%%%%%%%%%%%%%%%%%%%%%%%%%%%%%%%%%%%%%%%%%
%%%%%%%%%%%%%%%%%%%%%%%%%%%%%%%%%%%%%%%%%%%%%%%%

\section{\label{sec:conclusion}Conclusions}

We extended previous Doppler + self-lensing models to include eccentric orbits,
motivated by searches for sub-pc separation SMBHBs via unique periodic
signatures in their continuum light curves caused by the relativistic Doppler
boost and gravitational lensing of an accreting binary. We used these models
to investigate the optical light curves of two intriguing quasars, Ark 120 and
KIC 11606854 (Spikey).

The sinusoid-like light curve of Ark 120 suggests a binary candidate with a
20-yr period; two prominent flares suggest an eccentric orbit with lensing.
We find no evidence that a SMBHB Doppler + self-lensing model can describe the flares observed in the Ark 120 light curve. While we do not rule out a Doppler boost only model for Ark 120, it is disfavoured as our models predict a binary mass that is two orders of magnitude larger than the central mass estimate from single-epoch broad-line measurements.

The light curve of Spikey appears to be non-sinusoidal if periodic and has a
narrow symmetric spike, suggesting an eccentric orbit and lensing. We fit our
Doppler + self-lensing model to the data and find parameters that suggest a
total binary mass of $M_{\text{tot}} \approx 3 \times \sim 10^{7} \Msun$ and
rest-frame orbital period $T=418$ days. We find that the combination of
Doppler + self-lensing + DRW model provides a better fit for the variability
than the DRW model alone. This interpretation can be tested by monitoring
Spikey for periodic behavior and recurring spikes, the next of which are set
to occur in April 2020 and July 2022 (Figures \ref{fig:Spikeypredict} and
\ref{fig:Alldat}).

Because future searches for flares may be even cleaner in X-rays, since X-ray
emission is more compact and can be magnified by a larger factor, we have
obtained the first X-ray data on Spikey using the Swift observatory. Though
not taken during a predicted flaring period, these data show that Spikey is a
bright source of X-rays; hence, future X-ray observations have the opportunity
to detect the next lensing flare predicted here. The lack of a flare within
the predicted windows would rule out the SMBHB self-lensing hypothesis while
the detection of a repeating symmetric flare would be the most definitive
evidence to date for a sub-pc separation supermassive black hole binary.

\section*{Acknowledgements}
The authors thank Alberto Sesana, Thomas Kupfer, and Matthew Graham for useful
discussions. DJD acknowledges support from NASA through Einstein Postdoctoral
Fellowship award number PF6-170151 and funding from the Institute for Theory
and Computation Fellowship. ZH acknowledges support from NSF grant 1715661 and
NASA grants NNX17AL82G and 80NSSC19K0149. KLS acknowledges support from
Einstein Postdoctoral Fellowship award number PF7-180168. MC acknowledges
support from the National Science Foundation (NSF) NANOGrav Physics Frontier
Center, award number 1430284. Funding for this work was partially provided
through Swift proposal \#1518206. This paper includes data collected by the
Kepler mission. Funding for the Kepler mission is provided by the NASA Science
Mission directorate.

\bibliographystyle{mnras}
\bibliography{SMBHBs}

\section{Appendix}

%%%%%%%%%%%%%%%%%%%%%%%%%%%%%%%%%%%%%%%%%%%%%%%%
%%% FIGURE: CORNER PLOT %%%
%%%%%%%%%%%%%%%%%%%%%%%%%%%%%%%%%%%%%%%%%%%%%%%%
\begin{figure*}
\begin{center}$
\begin{array}{c}
\includegraphics[scale=0.3]{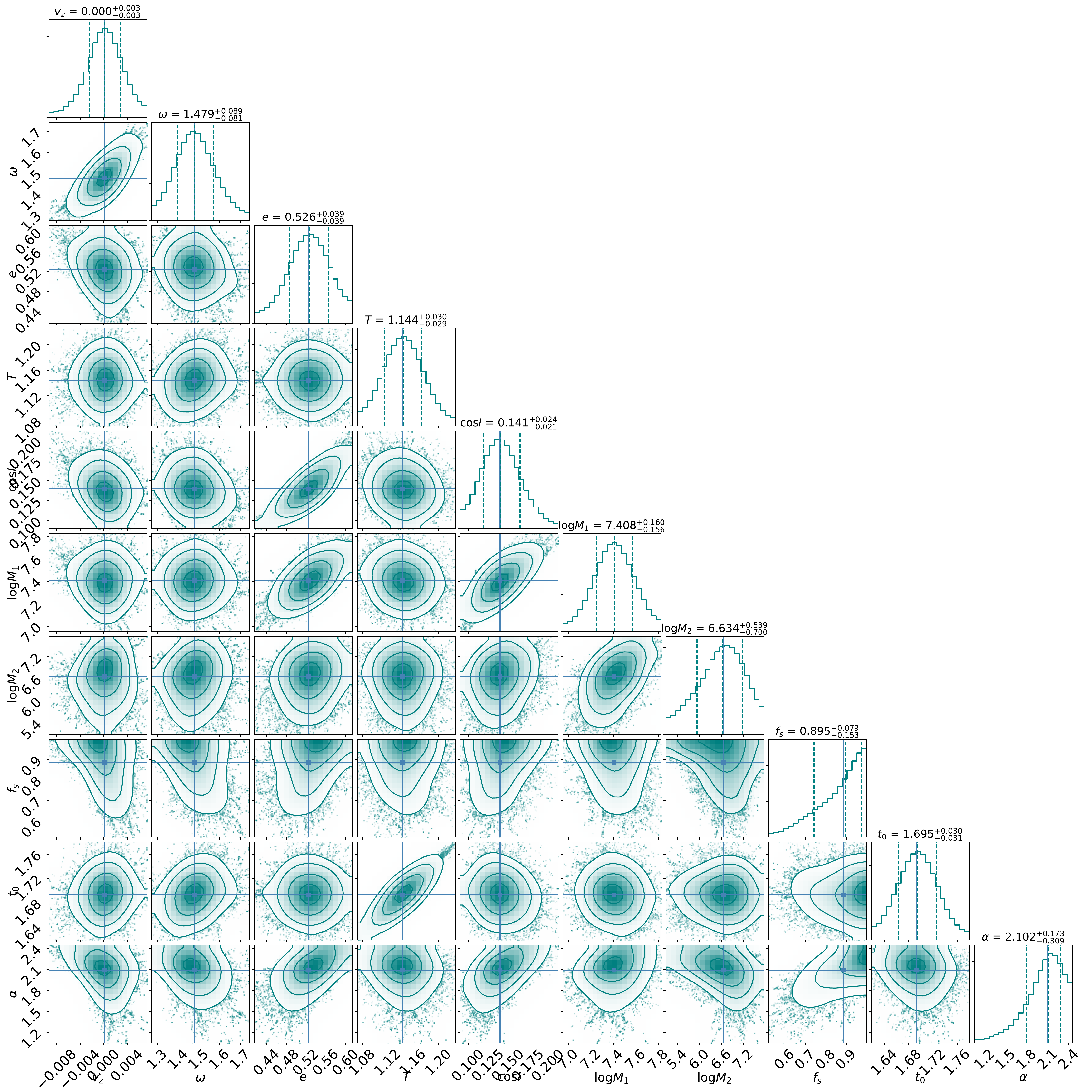}
\end{array}$
\vspace{-10pt}
\end{center}
\caption{\label{fig:agn_corner_500_2000}
One and two dimensional projections of the posterior probability distributions of the parameters, for Spikey. The intersecting lines on the 2D-posterier contours signify the $50\%$ likelihood values of the parameters, while the quoted parameters above the 1D distributions are uncertainties derived from the $50\%$ $16\%$, and $84\%$ quantile values using the last 100 steps for each walker. Compare with the maximum likelihood values listed in Table \ref{tab:parameters}, along with the relevant units. Corner plot made using \textit{corner} \citep{corner}. }
\end{figure*}
%%%%%%%%%%%%%%%%%%%%%%%%%%%%%%%%%%%%%%%%%%%%%%%%

\subsection{MCMC statistics}
\label{A:MCMC}
We include the assumed prior range of values in Table \ref{tab:parameters}. For completeness we provide the one and two dimensional posterior distributions sampled by emcee \citep{foremanmackey2013} for our more important result, Spikey. A degeneracy between parameters $M$ and $I$ already exists for circular orbits. When we take into account eccentric orbits, we see additional degeneracies between $e$ and $M$, and $e$ and $I$, as discussed in \S \ref{sec:model}. In the present case, the range of parameter values over which these degeneracies are present is small.

\end{document}